\documentclass{aa}
\usepackage{graphicx}
\usepackage{txfonts}
\usepackage{natbib}
\bibpunct{(}{)}{;}{a}{}{,}

\begin{document}

\title{Structure in the motions of the fastest halo stars}
\author{P. Re Fiorentin \inst{1,2}
	\and
	A. Helmi \inst{3}
	\and
	M.G. Lattanzi \inst{2}
	\and
	A. Spagna \inst{2}
	}
\offprints{P. Re Fiorentin, \email{paola.re.fiorentin@to.infn.it}}
	
\institute{Dipartimento di Fisica Generale, Via Pietro Giuria 1, 10125 Torino, Italy.
	\and 
	INAF - Osservatorio Astronomico di Torino, Strada dell'Osservatorio 20, 10025 Pino Torinese, Italy.
	\and
	Kapteyn Astronomical Institute, P.O. BOX 800, 9700 AV Groningen, 
The Netherlands.}

\date{Received ...., 2005; accepted ..., 2005}

\abstract{ We have analyzed the catalog of $2106$ non-kinematically
selected metal poor stars in the solar neighborhood published by
\citet{pap2}, with the goal of quantifying the amount of substructure
in the motions of the fastest halo stars. We have computed the
two-point velocity correlation function for a subsample of halo stars
within $1-2$~kpc of the Sun, and found statistical evidence of
substructure, with a similar amplitude to that predicted by high
resolution CDM simulations. The signal is due to a small kinematic
group whose dynamical properties are compared to the stellar "stream",
previously discovered by \citet{AHnat}.  If real, this high velocity
moving group would provide further support to the idea that
substructures remain as fossils from the formation of the Galaxy as
expected in the CDM scenario.  
 } 

\maketitle

\section{Introduction}
Observations of metal-poor stars have many times been used to 
discriminate (although not yet conclusively) between alternative
galaxy formation scenarios \citep[e. g., ][]{ELS,
SZ}. In recent years, considerable effort has been put into
understanding the properties of galaxies within the hierarchical
paradigm of structure formation in the Universe which, so far, looks
to be the most successful theory. In a CDM Universe, the first objects
to form are small galaxies which then merge and give rise to the
larger scale structures we observe today.  Thus, structure formation
occurs in a `bottom up' fashion.  This theory predicts the presence of
substructures (tidal tails, streams) due to the mergers and accretion
that galaxies have experienced over their lifetime.

Direct comparisons to observations have shown that this model can
reproduce the properties of both the local and the distant Universe.
Several examples of mergers and galaxy interactions have been observed
in the Milky Way, such as the disrupted Sagittarius and Canis Major
dwarf galaxies \citep[e. g., ][]{ibata, martin}, the phase-space
stream of halo stars in the solar neighborhood \citep{AHnat}, and the
ring in the outer Galaxy \citep{newberg}. Similar substructures have
also been found in the halos of other nearby galaxies, such as M31
\citep{ferguson} and NGC 5907 \citep{NGC5907}, showing that accretion
may be a common phenomenon in the evolution of galaxies.

Although the stellar halo accounts for only about $1\%$ of the
luminous mass, it plays a crucial role in studies of formation and
evolution of the Galaxy.  Signatures of the hierarchical nature of
galaxy assembly are expected to be most obvious in this
component. Moreover, stars in the halo are generally very old and
metal poor, i.e. they can be considered more pristine. These
are in fact the stars thought to have been formed in satellite
galaxies that merged to form our Galactic halo (Robertson et al. 2005).

At present, the best measurements of the halo kinematics are obtained
from the analysis of samples of stars located in the solar
neighborhood. Studies of the kinematics of various stellar populations
in the Galaxy have long been limited - especially for the inner Halo -
by the lack of large samples of stars with accurate distances,
metallicity and kinematics.

\citet{pap2} compiled an extensive catalog of metal-poor stars
selected without kinematic bias, and with available proper motions,
radial velocities, and distance estimates for stars with a wide range
of metal abundances. In this paper we analyze this data-set, which has
already provided support for constraining plausible scenarios for the
formation and evolution of the Milky Way 
\citep[e. g., ][]{pap3, CB2001}.

The layout of this paper is as follows: In Sect.~2, we assemble a
sample of metal-deficient ($\mathrm{[Fe/H]\le -1.5~dex}$) halo stars
up to $2$~kpc of the Sun selected from the \citet{pap2} catalog.  In
Sect.~3, we explore their phase-space distribution and quantify
clustering by means of the two-point correlation analysis. In Sect.~4,
we compare to theoretical predictions from CDM simulations. In Sect.~5,
we further examine evidence for kinematic substructure in the space of
adiabatic invariants.  The conclusions of the present study are given
in Sect.~6.

\section{Halo samples from Beers et al. (2000) catalog}

Beers et al. (2000) presented a revised large catalog of $2106$ metal
poor stars in the solar neighborhood selected without kinematic bias.
Within this sample, $1258$ stars have distance estimates and radial
velocities as well as proper motions, so that the three components of
the space velocities can be derived.

Throughout this work we adopt Galactic velocity components relative to
the Galactic center, where as usual $U, V$ and $W$ are positive toward
the center, in the direction of the rotation, and toward the north
Galactic pole, respectively.  We assume that the local standard of
rest (LSR) rotates with a velocity of $220~\mathrm{km~s^{-1}}$ about
the Galactic center, and that the peculiar velocity of the Sun
relative to the LSR is given by $(10.00,5.25,7.17)~\mathrm{km~s^{-1}}$
\citep{sun_lsr}.

\subsection{Selection criteria and selected sets}

Since we are interested in constructing a sample of halo stars, we
will only consider those stars in the Beers et al. (2000) catalog
with metallicities $\mathrm{[Fe/H]} \le -1.5$~dex.

To minimize any possible contamination from the thick disk, we further
exclude low metallicity stars near the plane and with coplanar and
circular orbits, i.e. with disk-like kinematics.  Assuming an
exponential thick disk with scale height
$0.8~\mathrm{kpc}<h_Z<1.5~\mathrm{kpc}$ \citep{hZm,hZM} and with a
velocity ellipsoid $(\sigma_R, \sigma_{\phi}, \sigma_Z; \langle
V_{\phi}\rangle)= (61, 58, 39; -36)~\mathrm{km~s^{-1}}$ \citep{GA}, we
exclude those stars for which the following conditions are
\emph{simultaneously} satisfied: $\vert Z \vert \le 1.5$~kpc,
$126~\mathrm{km~s^{-1}}\le V_{\phi}\le 242~\mathrm{km~s^{-1}}$, $\vert
V_R\vert \le 61~\mathrm{km~s^{-1}}$, low angular momentum
$\mathrm{L_{xy}=\sqrt{L_x^2+L_y^2}}$ with respect to $\mathrm{L_z}$,
namely $\mathrm{|{\bf L}| \le 1.1~L_z}$.

\begin{table}[htb]
  \caption{Selected sets of metal deficient
        ($\mathrm{[Fe/H] \le -1.5~ dex}$) halo stars. See text for
        explanation.}
  \label{sets}
  \begin{center}
    \leavevmode
        \begin{tabular}[h]{llcc}
        \hline \\[-5pt]
        Type & & $1$~kpc& $2$~kpc \\[+5pt]
        \hline \\[-5pt]
        D  & main-sequence dwarf star      & $11$ & $13$  \\
        A  & main-sequence A-type star     & $2$  & $3$   \\
        TO & main-sequence turnoff star    & $73$ & $169$ \\
	SG & subgiant star                 & $1$  & $9$   \\
	G  & giant star                    & $78$ & $133$ \\
	AGB& asymptotic giant branch star  & $5$  & $5$   \\
	FHB& field horizontal-branch star  & $5$  & $5$   \\
	RRV& RR Lyrae variables            & $16$ & $70$  \\
	V  & variable star                 & $1$  & $3$   \\
       \hline \\[-5pt]
        TOTAL&                             & $192$& $410$ \\[+5pt]
       \hline
       \end{tabular}
  \end{center}
\end{table}

The remaining stars define our \emph{selected} sample. This sample
contains $192$ stars within $1$~kpc and $410$ within $2$~kpc of the
Sun. These sets include subdwarfs, giants, and variables, as described
in Table \ref{sets}, with the following properties:
\begin{itemize}
\item proper motions are available from the HIPPARCOS Catalog for $197$ 
stars; the other ground-based measurements come from 
the SPM Catalog 2.0, Lick NPM1 Catalog, STARNET Catalog, 
ACT Reference Catalog; for stars independently measured in two or more 
catalogs, we have averaged the proper motions weighted by their errors; 
in all cases accuracies of a few $\mathrm{mas~yr^{-1}}$ are achieved.
\item Radial velocities have typical accuracies of the order of 
$10~\mathrm{km~s^{-1}}$. Metal abundances have been determined either 
spectroscopically or from suitable photometric calibrations.
\item Calibrations of absolute magnitude $M_V$ allow photometric 
parallaxes for which Beers et al. (2000) claim an accuracy of 
$\sim  20\%$. 
Because of the large distance ($d>200$~pc), these photometric parallaxes 
are, formally, more precise than the corresponding HIPPARCOS trigonometric 
parallaxes, however they still remain the main source of uncertainty in the 
derived tangential velocities.
\end{itemize}

\section{Kinematic analysis}

\begin{figure*} 
   \centering
   \includegraphics[angle=0,width=17cm]{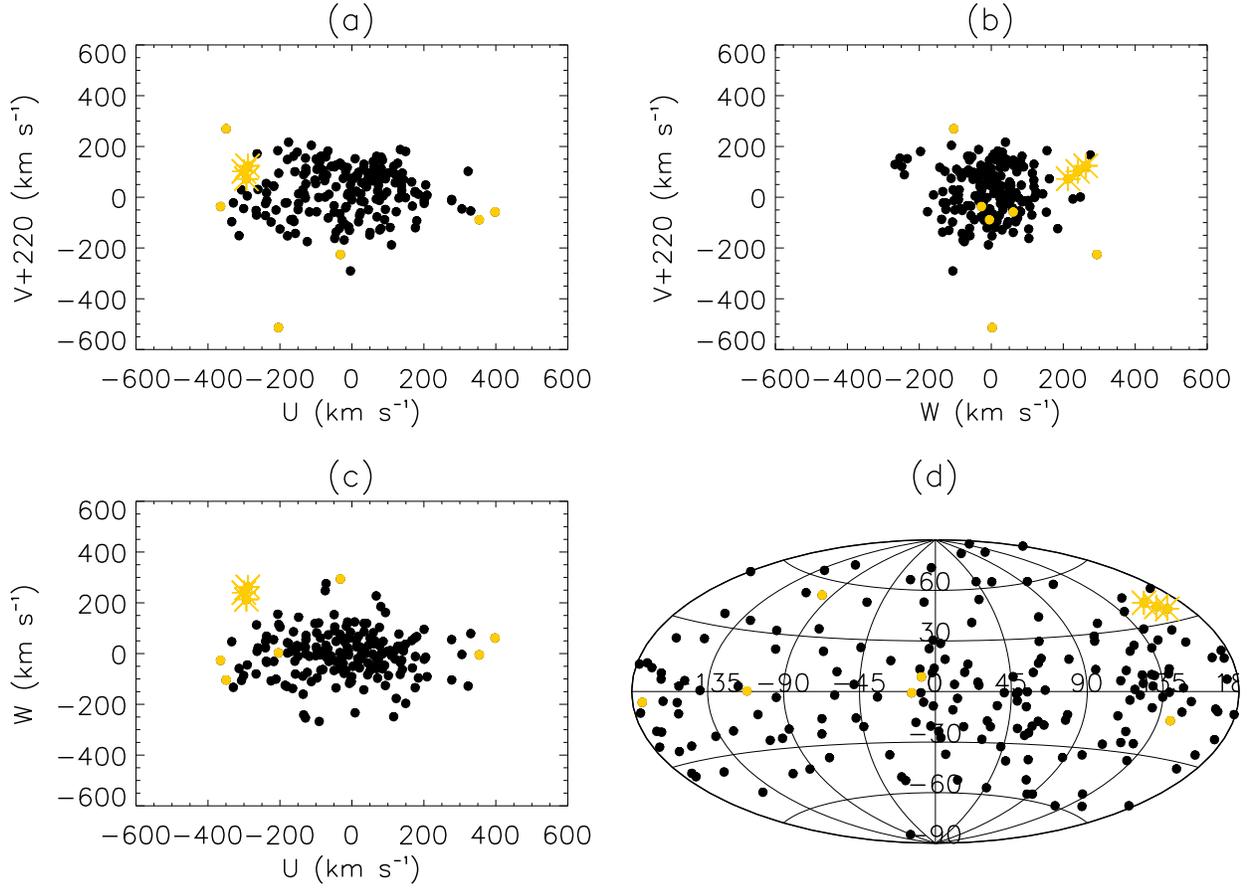}
   \caption{Distribution of nearby halo stars in velocity space for our
        selected sample with $\mathrm{[Fe/H]\le -1.5}$~dex within
        $1$~kpc of the Sun.
        (a-c) Velocity projections in the $(U,V,W)$ space;
        (d) Velocity direction: the position of each particle is given by the
        spherical angular coordinates
        ${\mbox{\boldmath $\alpha$}}=(\phi, \theta)$ of the velocity vector.
        Of the $192$ stars present in this volume, the $5\%$ fastest are
        highlighted as light color dots. Among them, 
	the asterisks identify
        the objects/groups with velocity difference less than
        $\mathrm{42~km~s^{-1}}$.}
   \label{fig_UVW}
\end{figure*}

Analytic arguments and high-resolution cosmological simulations of the
formation of dark matter halos (Helmi \& White 1999; Helmi et
al. 2002; Moore et al. 2001) suggest that the halo should be spatially
smooth in the Solar neighborhood (although see some recent discussion
in Diemand et al. 2005; and Zhao et al. 2005). This is also supported
by measurements of the degree of lumpiness in the angular distribution
of halo stars (e.g. Lemon et al. 2004). However, the same simulations
predict that while the kinematics of halo objects near the Sun can be
represented by a smooth multivariate Gaussian distribution, the
motions of the most energetic particles should be strongly clumped and
anisotropic.  This result has motivated us to analyze in more detail
the kinematics of the halo samples we identified above.

\subsection{Velocity distribution}

The motion of each star in the sample can be specified by the velocity
vector ${\bf v}=(U,V,W)$ in 3D linear space, as well as in the 2D
angular space, by spherical angular coordinates
${\mathrm{\mbox{\boldmath $\alpha$}}}=(\phi, \theta)$, where
$\sin{\theta} = V_z/|{\bf v}|$ and $\tan\phi = V_y/V_x$.

Figure~\ref{fig_UVW} shows the kinematic distribution of the selected
sample within $1$~kpc.  In Fig.~\ref{fig_UVW}(a-c) we plot the $U,V,W$
velocities. Fig.~\ref{fig_UVW}(d) shows the distribution of velocity
directions $\mbox{\boldmath $\alpha$}=(\phi, \theta)$.

The velocity distribution is relatively smooth, and appears to be
consistent with a Gaussian Schwarzschild distribution. The mean velocities 
are $(\langle U\rangle,\langle V+220\rangle,\langle W\rangle)=
(-26\pm 11,21\pm 8,-1\pm 7)~\mathrm{km~s^{-1}}$ and the  velocity 
ellipsoid is radially elongated, namely
$(\sigma_U,\sigma_V,\sigma_W)=(150\pm 8,106\pm 5,96\pm 5)~\rm{km~s^{-1}}$. 
However, a smooth description does not seem to reproduce the kinematics of the
the fastest objects, as shown by the highlighted points.  The 5\%
fastest moving stars (light color dots) seem to be more clumped.
The characteristics of the observed velocity distribution appear to be
in rough agreement with the results of CDM simulations as we shall
show in Sect.5. 

We shall perform two different statistical tests on the data with the
aim of quantifying the presence of large and small scale anisotropies
in the motions of our stars. To establish the significance of our
results we will compare our results to suitable Monte Carlo
simulations. 

Our synthetic data sets have the same number of stars and the same
spatial distribution as the observed sample. The characteristic
parameters of the multivariate Gaussian used to describe the
kinematics are obtained by fitting to the observed mean values and
variances after appropriate convolution with observational errors. We
generate 100 `observed' samples as follows.  A velocity is drawn from
the underlying multivariate Gaussian; it is transformed to a proper
motion and radial velocity (assuming the observed parallax and
position on the sky); observational `errors' of the magnitude
described in Sect.2 are added to the parallax, the radial velocity and
the proper motion; these `observed' quantities are then transformed
back to an `observed' velocity. 

The first test we perform on the data consists in quantifying the
presence of large-scale anisotropies. We implement this test by
partitioning the 2D angular space $(\phi,\theta)$ -see
Fig.~\ref{fig_UVW}(d)- into cells with roughly similar area.  We then
count how many stars fall in each cell and compare it to the expected
number in our Monte Carlo simulations.
 
\begin{figure}
   \centering
   \resizebox{\hsize}{!}{\includegraphics[angle=90,width=9cm]
        {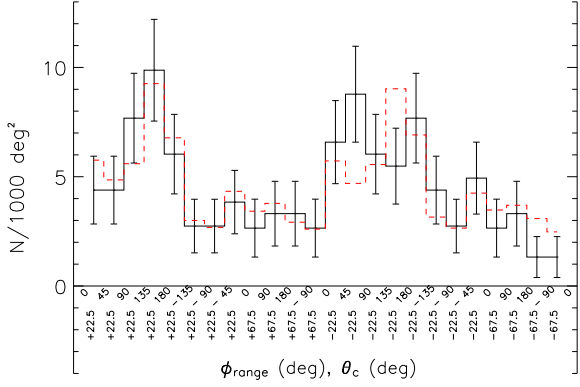}}
   \caption{$(\phi,\theta)$-space number counts for a partition with
     $24$ cells with $\Delta\theta=45^\circ$ (mean value $\theta_{\rm
     c}$) and $\Delta\phi=45^\circ/90^\circ$ for the stripes along the
     equator and close the poles respectively.  The solid line
     corresponds to the selected sample within $1$~kpc of the Sun,
     while the dashed histogram to the average number of counts in our
     Monte Carlo simulations. Error bars are based on Poisson
     statistics.  }
   \label{anisotropy}
\end{figure}

Figure~\ref{anisotropy} shows the results for the partition with $24$
cells (with an area of $1800-1500~{\rm deg^2}$) along $4$ strips:
$0^\circ<|\theta|<45^\circ$ (with $\Delta\phi=45^\circ$) and
$45^\circ<|\theta|<90^\circ$ (with $\Delta\phi=90^\circ$).  Plotted
are the counts in each cell for the selected halo sample within
$1$~kpc of the Sun (solid line) and for the average of $100$ Monte
Carlo realizations (dashed line).  The east-west large-scale
anisotropy seen in this figure is due to the radially elongated
velocity ellipsoid in combination with a small amount of prograde
rotation.

We find, at the $1~\sigma$ level, an excess of stars due to residual
contamination from the thick-disk ($\phi\sim 90^\circ$) and a lack of
stars moving toward the SGP ($\theta\sim -90^\circ$).  To this regard, it
is interesting to note that this deficiency is at odds with some
models of the Sagittarius dwarf evolution that predict the presence of
a stellar stream that should cross the Solar neighborhood. The stream
would be visible as an excess of stars moving toward the SGP
(e.g. Helmi 2004a; Law et al. 2005). Even though it could be argued
that this sample may be too bright and possibly too metal-poor, this
result could rule out an oblate or spherical shape for the Galactic
dark-matter halo, and favor a prolate mass distribution as suggested
by Helmi (2004b).

In summary, the observed and simulated counts are statistically
indistinguishable for all partition choices. There is no evidence
(at $>2\sigma$-level) for large scale flows crossing the Solar neighborhood.

\subsection{Correlation function}

We quantify the deviations from a smooth Gaussian distribution due 
to kinematic substructures by means of the two-point correlation 
function, defined as
\begin{equation}
\xi =\frac{\langle DD\rangle}{\langle RR\rangle} -1
\end{equation}
where $\langle DD\rangle$ is the number of pairs of stars in our data
with velocity difference less than a given value, namely
\begin{equation}
\langle DD\rangle =\sum {{\rm pairs~of~stars}~i,j~{\rm with}~\vert
{\bf v}_i-{\bf v}_j \vert \le \Delta}
\end{equation}
and $\langle RR\rangle$ is defined analogously for the same number of
random points drawn from a multivariate Gaussian distribution derived
from the data set, convolved with expected observational errors, and
averaged over a hundred realizations. Note that the points are not
statistically independent, and that our definition of $\xi$ actually
corresponds to the cumulative correlation function often used in
cosmology in studies of the large-scale structure of the Universe
\citep[e.g., ][]{coil, mullis}.

Based on Poisson counts, we estimate the error of the two-point 
correlation function as 
\begin{equation}
\Delta_{\xi}=\frac{1+\xi}{\sqrt{\langle DD\rangle}} 
\end{equation}
  
\begin{figure}
   \centering
   \resizebox{\hsize}{!}{\includegraphics[angle=90,width=9cm]
        {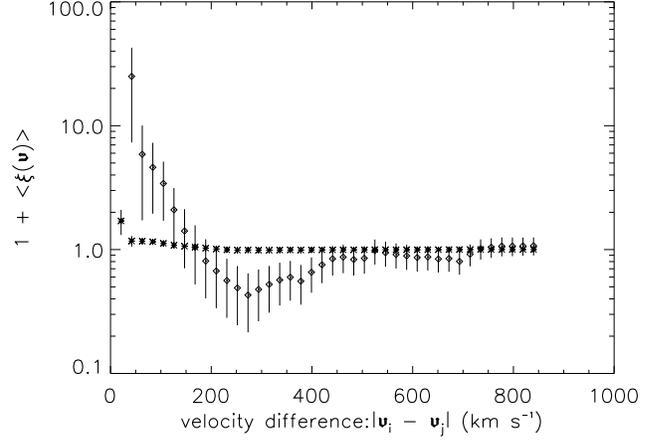}}
   \caption{Two-point velocity correlation function for our selected
     sample within 1 kpc (asterisks) and for the 5\% fastest moving stars
     (diamonds).  In both cases, there is a signal at small velocity
     differences, indicative of the presence of streams. The error
     bars are based on Poisson counts.}
   \label{fig_CF}
\end{figure}

Thus, in the 3D linear space $(U,V,W)$ - say $v=\vert {\bf v}_i-{\bf
v}_j\vert$ - $\xi(v)$ measures the excess of pairs of stars moving
with velocity difference equal or smaller than a given value, above
that expected from a random sample. Clumping due to kinematic
substructures (i.e. groups of stars moving with similar velocities) is
manifested by an excess at small velocity separations.

In Figure~\ref{fig_CF} we show the two-point correlation function
$\xi(v)$ for our selected sample within $1$~kpc of the Sun (asterisks)
and a subset comprising the $5\%$ fastest-moving stars (diamonds).  We
use bins of width $21~\rm{km~s^{-1}}$, up to separations of
$840~\rm{km~s^{-1}}$.

The figure shows that, at a $1~\sigma$ level, there is a small but 
statistically significant excess of stars with similar velocities with 
respect to what would be expected for a smooth Gaussian distribution. 
The signal in the first bin is weak but significant.
No correlation is observed at larger separations.

However the signal is clearly much stronger for the subset of $5\%$
fastest moving stars: in this case, the excess of pairs of stars with
similar velocities is very noticeable, and it is a direct indication of
the presence of clumps/streams. This signal is indeed due to a moving
group, which is formed by three stars (described in
Table~\ref{pstream}) and indicated by the asterisks in
Fig.~\ref{fig_UVW}.

\begin{table*}
      \caption[]{Members of the identified moving group.}
        \hspace{21 cm}
         \label{pstream}
     $$
         \begin{array}{lccccccccc}
            \hline
            \noalign{\smallskip}
            \mathrm{Name} & \mathrm{[Fe/H]} & \alpha~\mathrm{(J2000)}
                & \delta~\mathrm{(J2000)} & D & U & V+220 & W \\
              & \mathrm{(dex)} & \mathrm{(h~m~s)} & \mathrm{(d~m~s)}
                & \mathrm{(kpc)}& \mathrm{(km~s^{-1})}
                & \mathrm{(km~s^{-1})}& \mathrm{(km~s^{-1})}\\
            \noalign{\smallskip}
            \hline
            \noalign{\smallskip}
            \mathrm{BPS~CS~30339-0037} & -2.13  &  00~20~28.90
                & -36~12~00.7  &  0.80\pm 0.16 & -292\pm 54 &  73\pm 36 & 214\pm 12\\
            \mathrm{HD~214161}           & -2.16  &  22~37~08.04
                & -40~30~38.4  &  0.62\pm 0.12 & -290\pm 23 & 123\pm 22 & 262\pm 15\\
            \mathrm{V^{*}~RZ~Cep}      & -1.77  &  22~39~13.05
                & +64~51~28.9  &  0.41\pm 0.08 & -301\pm 62 & 101\pm 27 & 240\pm 47\\
            \noalign{\smallskip}
            \hline
         \end{array}
     $$
\end{table*}

Note also the presence of a certain degree of anti-correlation for the
interval $200-400~\mathrm{km~s^{-1}}$.  This may be due to a type of
`clear out effect' as the result of the clumping of few objects within
the first bins. This effect would have a considerably smaller
amplitude for a larger sample (as is the case when our selected sample
is considered in its entirety).

So far we have focused on the $5\%$ fastest moving particles within
$1$~kpc of the Sun.  Analysis performed with the $10\%$ fastest
subsample still shows a significant deviation from a multivariate
Gaussian, albeit of smaller amplitude.

Turning to an eight times larger volume (we consider those stars
within $2$~kpc of the Sun) allows us to increase the number of stars,
but only by a factor of $\sim 2$. This increase in the number of
stars, does not necessarily translate into an enlargement of the
number of streams but could also lead to a better representation of
each stream.

However this effect is not obvious in our sample: for the larger
volume the kinematic group previously singled out loses one star, 
BPS CS 30339-0037, 
which is no longer selected in the $5\%$ high velocity 
tail (simply a consequence of small number statistics).  Up to
velocity difference smaller than $42~\mathrm{km~s^{-1}}$, a weak
signal (at the $1~\sigma$ level) is noticeable: it is due to the
two remaining stars of our moving group and a second structure with
two other stars. Given the low amplitude of this signal in comparison
to the 1~kpc sample, we are led to believe that this second structure
is probably a statistical fluke rather than a true physical system.

In summary, we have detected a moving group (with three members) among
the 5\% most energetic stars within $1$~kpc of the Sun; no other
members could be found by increasing the volume or relaxing the
velocity threshold (from 5 to 10\%).

\section{Comparison to the Theory: Simulations}

\begin{figure*}  
   \centering
   \includegraphics[angle=0,width=17cm]{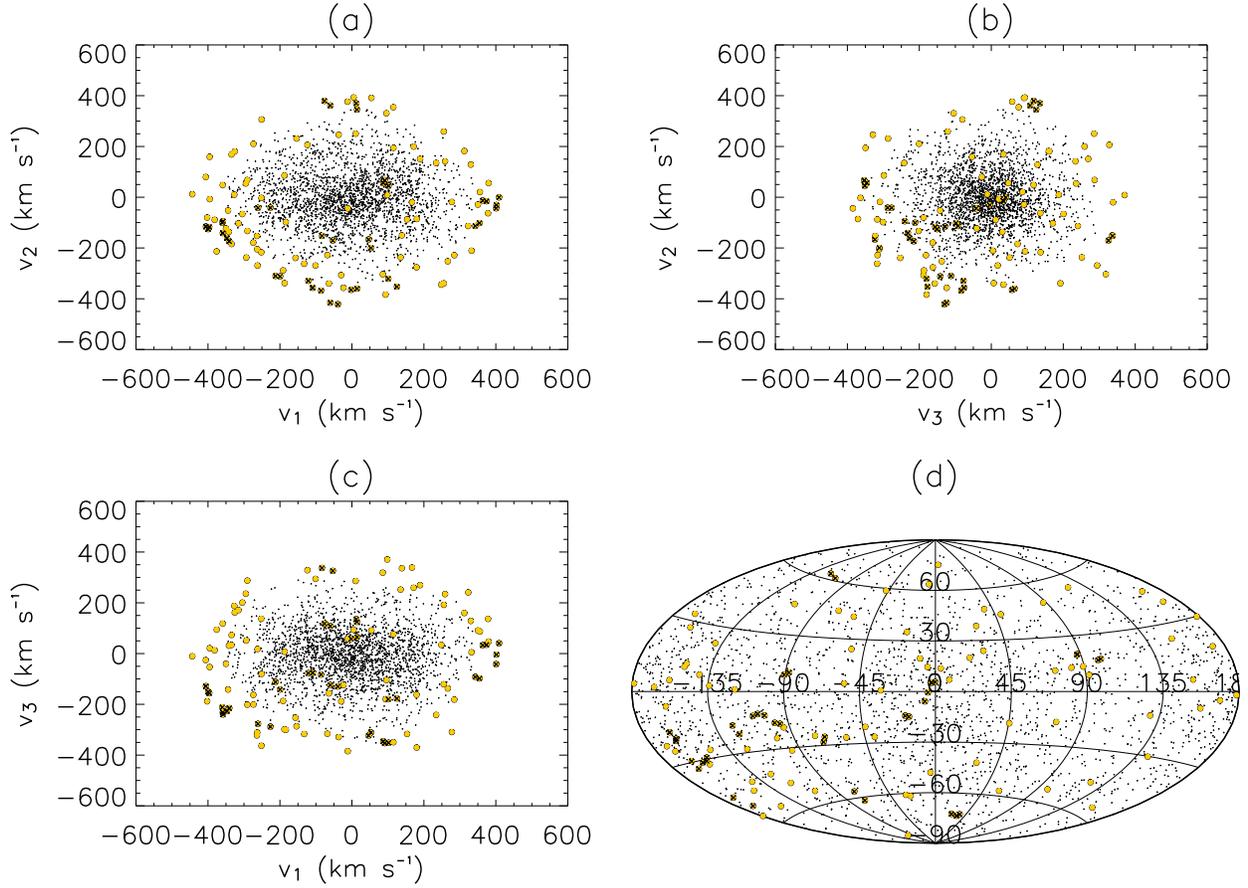}
   \caption{Distribution of 2348 dark matter particles from a CDM
   simulation located in a sphere of 2~kpc radius centered at $8$~kpc
   from the Galaxy center. (a-c) Velocity projections in the
   $(v_1,v_2,v_3)$ space; (d) Velocity directions. The solid gray
   circles denote the 5\% most energetic particles, while the
   asterisks identify those subsets whose velocity difference is less
   than $42~\mathrm{km~s^{-1}}$. The kinematics characteristics
   observed in this set of plots are representative of what is seen in
   other similar volumes.}
   \label{SIM}
\end{figure*}

We now wish to compare our results from the previous section to
theoretical models. To this end, we use a high-resolution simulation
of the formation of a dark matter halo in a $\mathrm{\Lambda CDM}$
cosmology (Springel et al. 2001).  From this simulation, we derive the
kinematics of dark matter particles inside spheres of $2$~kpc radius,
located at $8$~kpc from the Galactic center.  In doing so, we are
assuming that these volumes are representative of the solar
neighborhood.

\begin{figure} 
   \centering
   \resizebox{\hsize}{!}{\includegraphics[angle=90,width=9cm]
	{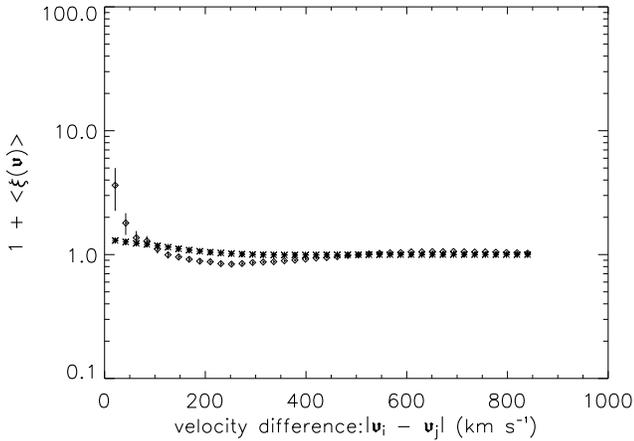}}
   \caption{Velocity correlation function $\xi(v)$ for the volume
     shown in Fig.~\ref{SIM}.  Asterisks and diamonds correspond to
     $\xi$ for the full sample and for the subset of 5\% fastest
     moving particles, respectively.}
   \label{CF_SIM}
\end{figure}

In Fig.~\ref{SIM} we plot the kinematics of $2348$ particles
selected within one of the $2$~kpc spheres.  Their velocity
distribution is relatively smooth and appears to be quite consistent
with a multivariate Gaussian (see Fig.~\ref{CF_SIM}), with principal
axes $(\sigma_1,\sigma_2,\sigma_3)=(143,118,110)~\mathrm{km~s^{-1}}$.
However, if we focus on the motions of the most energetic particles
(indicated with gray symbols in Fig.~\ref{SIM}) this is no longer
the case.  The 5\% fastest moving particles are strongly clumped, and
their distribution is highly anisotropic. 

To quantify the substructures present in this volume we compute the
velocity correlation function $\xi$ (as described in Sect.~3). The
random comparison sample in this case is the result of averaging 100
realizations of a trivariate Gaussian with similar moments as the dark
matter velocity distribution. In Fig.~\ref{CF_SIM} we observe a weak
signal in the first bins produced by a small excess of particles with
similar velocities (asterisks). However, if we focus on the 5\%
fastest moving particles (diamonds), the excess has a much larger
amplitude particularly at small velocity differences (i.e. $\Delta v <
42~\mathrm{km~s^{-1}}$), and simply reflects the presence of kinematic
groups clearly visible in Fig.~\ref{SIM}.

Although the number and the properties of stellar streams are likely
to be rather different from pure dark-matter streams, it is worth
noting that these results are qualitatively similar to those found for
our stellar samples. To try to quantify the degree of similarity, it
is simplest to assume that 10\% of the particles in these volumes
represent stars (i.e. reflecting a ``universal'' baryon fraction). In
this way we can randomly define ``stellar samples'' which we subject
to statistical analysis, such as the velocity correlation
function. Such analysis show a weak signal at the $1\sigma$ level,
which has a smaller amplitude than found for the selected stellar
sample discussed in Sect.~3. This result could suggest that stars are
not just a random subset of dark-matter particles. This would not be
very surprising since stars are expected to be much more clustered in
the centers of dark-matter halos.

In principle it should be possible to identify which particles might 
represent stars using better motivated physical arguments (e.g.,
by selecting those that have the largest binding energies at redshift
$\sim 10$, Moore 2001). However, this is not straightforward, as it
involves, for example determining the efficiency of star formation in
each progenitor halo (Robertson et al. 2005).

In any case, our main limitation in quantifying the degree of
similarity between the stellar sample and the dark-matter simulation
lies in the number of particles available in this simulation. We
expect that typically, only $1$ in $250$ particles will be a star
\citep{AH03}\footnote{This is essentially determined by the ratio of
mass in dark-matter to that in halo stars for the Solar
neighborhood.}.  If this is the case, to reproduce our selected sample
with $\sim 400$ stars our simulations should have $10^5$ particles in
each volume.

\section{Adiabatic invariants}

\begin{figure*}
   \centering
   \includegraphics[angle=0,width=17cm]{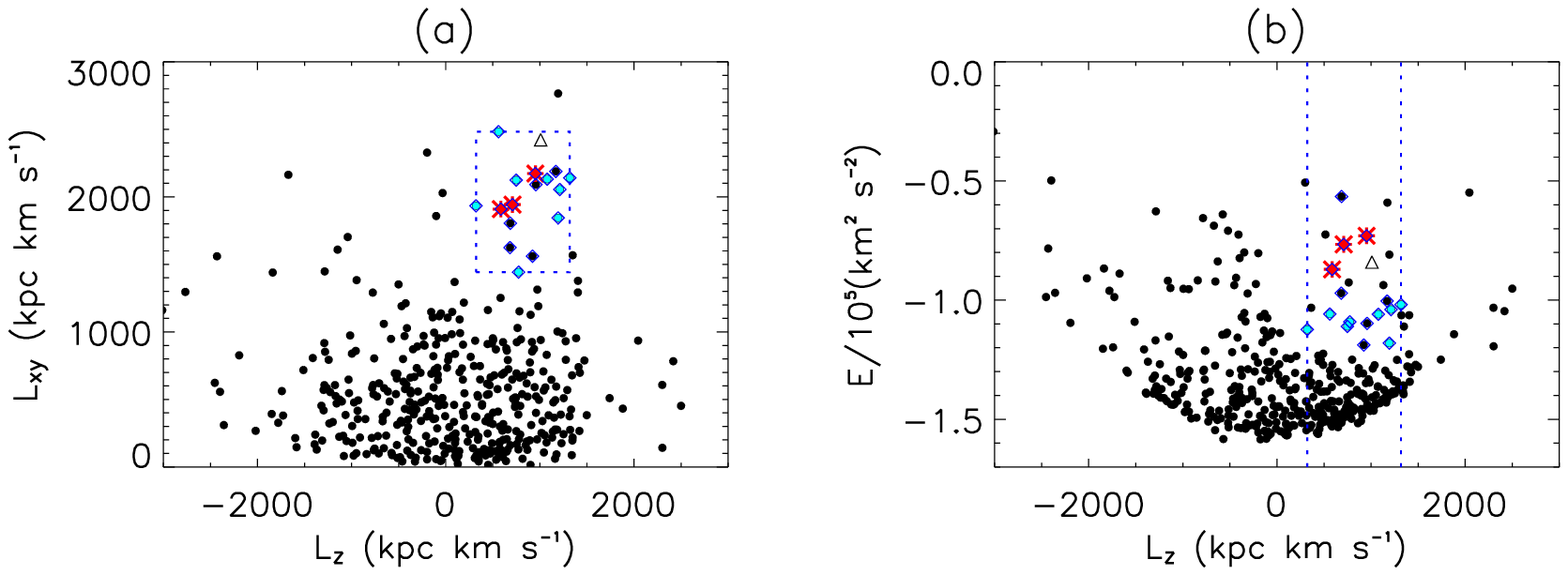}
   \caption{Distribution of the selected sample of $410$ stars with
     $D<2$~kpc in the space of adiabatic invariants.  The asterisks
     denote our kinematic group (RHLS), and the grey diamonds the
     clump identified by \citet{AHnat} (HWdZZ).  These stars are used
     to define a region (limited by the dashed lines in the left
     panel), which encompasses also other stars with similar angular
     momenta, and which are indicated as diamonds in both
     panels. $\mathrm{CD-80~328}$, the candidate from the Nordstr\"om
     et al. (2004) catalog, has been added and is shown as a triangle.
     }
   \label{FIG_AD}
\end{figure*}

We are also interested in understanding the properties of our selected
sample in the space of adiabatic invariants, since this is generally
the preferred space to look for evidence of substructures related to
past mergers.  Here clumping should be strong, since all stars
originating from the same progenitor have very similar integrals of
motion, resulting in the superposition of the corresponding streams.
In particular, \citet{AHnat} and \citet{pap3} have examined the
kinematics of metal-poor stars in the solar neighborhood and
identified a statistically significant clumping of stars in the
angular momentum diagram $L_z$ versus $L_{xy} =\sqrt{L_x^2+L_y^2}$.

The substructure identified by \citet{AHnat} consists of $7~(12)$
stars with $\mathrm{[Fe/H] \le -1.6~(-1.0)}$~dex and $D<1~(2.5)$~kpc,
and was discovered in a sample that was primarily the result of the
combination of HIPPARCOS data with the Beers \& Sommer-Larsen catalog
(1995). Chiba \& Beers, using the Beers et al. (2000) catalog (i.e. a
revised edition of the previous compilation) confirmed $9$ of their
stars in the clump region which includes also one of the stars in our
kinematic group, HD214161.  However, they later discarded this star
for having quite different orbital properties than other clump
members.

Here we focus on the space of adiabatic invariants $E$, $L_z$ and
$L_{xy}$, although the latter is not fully conserved in an
axisymmetric potential, and its use may not be generally appropriate
for the study of substructure in the halo. Nevertheless, we hope to
obtain further insight into our previous analysis based only on
kinematics.  We assume that the Galactic potential is represented by
three components \citep{JHB}: a dark halo with logarithmic potential,
a Miyamoto-Nagai disk, and a spherical Hernquist bulge:
\begin{equation}
$$\begin{array}{lll}
\Phi _{\rm halo}&=& v^2_{\rm h}\ln{\frac{r^2+d^2}{r^2_{200}}}\\
&&\\
\Phi_{\rm disk}&=&-\frac{GM_{\rm disk}}{\sqrt{R^2+(a+\sqrt{z^2+b^2})^2}}\\
&&\\
\Phi_{\rm bulge}&=& -\frac{GM_{\rm bulge}}{r+c}
\end{array}$$
\end{equation}
where $M_{\rm disk}=1.0\times 10^{11} M_{\sun}$, $M_{\rm bulge}=
3.4\times 10^{10} M_{\sun}$; $v_{\rm h}= 131.5~\mathrm{km~s^{-1}}$ and
$r_{200}=300~\mathrm{kpc}$; $a=6.5$~kpc, $b=0.26$~kpc, $c=0.7$~kpc,
and $d=12$~kpc.  This choice of parameters gives a circular velocity
at the solar radius of $210~\mathrm{km~s^{-1}}$.

Figure~\ref{FIG_AD} shows the distribution of the selected sample
within $2$~kpc of the Sun, in the space of adiabatic invariants.  Note
that the sample includes $8$ stars (hereafter HWdZZ stars) from the
stream identified by \citet{AHnat}. These stars are shown by gray
solid diamonds while our kinematic clump (hereafter RHLS) is plotted
as asterisks.  It is clear that these groups have similar momenta (see
left panel).  If $L_{xy}^{\rm min}$ ($L_z^{\rm min}$) and $L_{xy}^{\rm
max}$ ($L_z^{\rm max}$) denote the minimum and maximum $L_{xy}$
($L_z$) values for the HWdZZ and RHLS stars, the region $L_{xy}^{\rm
min}<L_{xy}<L_{xy}^{\rm max}$ and $L_z^{\rm min}<L_z<L_z^{\rm max}$
(dotted box) encompasses stars with similar momenta. There are $16$
stars located in this region, which include five new objects shown as
black diamonds in Fig.~\ref{FIG_AD}. Some of these stars are also
members of the `clump-trail' structure identified by \citet{pap3}
(hereafter CB).  The orbital properties of these 16 stars, based on
the Galactic potential defined above, are listed in
Table~\ref{tabAD}. The last column serves to indicate the membership
to the various groups.

\begin{table*}
  \caption{Characteristics of the stars located in the region
  $L_{xy}^{\rm min}<L_{xy}<L_{xy}^{\rm max}$ and $L_z^{\rm
  min}<L_z<L_z^{\rm max}$ defined by the Helmi et al. (1999) structure
  (within the box shown in the left panel of Fig.~\ref{FIG_AD}.)}
  \hspace{21 cm}
  \label{tabAD}
  \begin{center}
    \leavevmode
        \begin{tabular}[h]{lcccccr}
        \hline \\[-5pt]
        Name & $\rm{[Fe/H]}$ & $\rm{E}/10^5$ & $\rm{L_z}$ & $\rm{L_{xy}}$ &
	$e$&Membership\\
	     & (dex)& ($\rm{km^2~s^{-2}}$)& ($\rm{kpc~km~s^{-1}}$)& ($\rm{kpc~km~s^{-1}}$)&\\[+5pt]
        \hline \\[-5pt]
	BPS CS 22948-0093  &  -3.72 &  -0.57 &  687 & 1805& 0.92&   \\
	HD 214161          &  -2.16 &  -0.73 &  952 & 2173& 0.81&RHLS    \\
	V* RZ Cep          &  -1.77 &  -0.77 &  710 & 1943& 0.85&RHLS    \\
	BPS CS 30339-0037  &  -2.13 &  -0.87 &  586 & 1909& 0.81&RHLS    \\
	BPS CS 22189-0007  &  -2.12 &  -0.97 &  683 & 1624& 0.77&     \\
	BPS CS 29513-0031  &  -2.79 &  -1.00 & 1171 & 2189& 0.47&CB   \\
	BD+10 2495         &  -1.83 &  -1.02 & 1318 & 2140& 0.47&HWdZZ, CB\\
	HD 119516          &  -2.49 &  -1.04 & 1212 & 2055& 0.50&HWdZZ, CB\\
	V* TT Cnc          &  -1.57 &  -1.06 &  562 & 2483& 0.42&HWdZZ    \\
	CD-36 1052         &  -2.19 &  -1.06 & 1078 & 2132& 0.45&HWdZZ, CB\\
	HD 237846          &  -2.63 &  -1.09 &  774 & 1443& 0.71&HWdZZ    \\
	BPS CS 29504-0044  &  -2.04 &  -1.10 &  958 & 2091& 0.42&CB   \\
	V* AR Ser          &  -1.78 &  -1.12 &  322 & 1934& 0.46&HWdZZ    \\
	V* TT Lyn          &  -1.56 &  -1.11 &  748 & 2124& 0.45&HWdZZ, CB\\
        HD 128279          &  -2.20 &  -1.18 & 1194 & 1844& 0.23&HWdZZ, CB\\
	BPS CS 22876-0040  &  -2.20 &  -1.19 &  922 & 1561& 0.42&CB   \\[+5pt]
       \hline \\[-5pt]
       \end{tabular}
  \end{center}
  \begin{list}{}{}
\item [Membership:] RHLS: Re Fiorentin et~al. (this paper); HWdZZ: \citet{AHnat}; CB: \citet{pap3}
\end{list}    
\end{table*}

We have looked for additional members of our kinematic group in the
Nordstr\"om et al.  (2004) catalog of nearby stars.  Selecting stars with
$\mathrm{[Fe/H]\le -1.5~dex}$ and similar kinematic characteristics
as those we identified, we find only one possible
candidate. $\mathrm{CD-80~328}$ is a metal poor star
($\mathrm{[Fe/H]=-1.98~dex}$) with
$(U,V+220,W)=(-193,125,303)~\rm{km~s^{-1}}$ and a highly eccentric orbit
$e=0.83$.  It is shown as a triangle in Fig.~\ref{FIG_AD}.
$\mathrm{CD-80~328}$, having $L_z=1009~\rm{kpc~km~s^{-1}}$ and
$L_{xy}=2422~\rm{kpc~km~s^{-1}}$, is located well within the
box-region defined by the HWdZZ stars.  However, given its much lower
binding energy ($E=-0.84\cdot 10^5~\rm{km^2~s^{-2}}$), it is more
likely that $\mathrm{CD-80~328}$ is a member of our moving group.

According to numerical simulations carried out by \citet{brook}, stars
from an accreted satellite show typically highly eccentric orbits
($e>0.8$) and strongly correlated velocities. In view of this, the
stars of our kinematic group (with $0.81<e<0.85$) could well share a
common origin, and be stellar debris from an accreted
satellite.

It is worth noting that, while having similar momenta, our kinematic
structure and the HWdZZ clump have somewhat different energies.  This
could imply that these groups have different origin.  However, it
seems also plausible that these groups have been stripped off from the
{\it same} progenitor at different times.  In this scenario, HWdZZ
stars should have been released in a later galactic passage than those
in our kinematic clump. The difference in orbital energy between the
groups could be the result of different binding energies of the
progenitor galaxy (higher and lower, respectively) due to the effects
of dynamical friction upon this system while orbiting the Milky Way.

\section{Conclusions}
	We have extracted samples of metal-poor ($\mathrm{[Fe/H] \le
	-1.5}$~dex) halo stars in the Solar neighborhood from
	\citet{pap2} catalog, and carried out a statistical analysis
	of their kinematics.
 
	Based on clustering in the velocity space, we have found evidence of 
	substructures  in the motions of the
        fastest moving stars, at a level which
        seems to be consistent with that predicted by high resolution
        simulations of dark matter halos in a hierarchical universe.
	
	The moving group responsible for this signal is comprised by
	three stars, whose kinematic and metallicity characteristics
	are similar to the streams found by Helmi et al. (1999),
	albeit on somewhat more loosely bound orbits.

	Our sample of halo and high velocity stars is too small to
	make definite statements about the importance of accretion in
	the formation of the Galactic halo. This would require a
	sample with a few thousands nearby halo stars (i.e., $\sim 10$
	times larger than our selected sample) with accurate space
	velocities.
Such sample sizes will become available in the 
near future, thanks to spectroscopic surveys like RAVE (Steinmetz 2003) 
and SDSS-II/SEGUE (Beers et al. 2004), which could be combined 
with proper motion catalogs such as 
UCAC2 (Zacharias et al. 2004), SPM (Platais et al. 1998), 
GSC-II (McLean et al. 2000), USNO-B (Monet et al. 2003), to obtain 
full phase-space information. 

The existence of structures in the halo, if confirmed by
further studies, is of great importance for constraining
models of the formation and evolution of the Galaxy. The space
astrometric mission Gaia (Perryman et al. 2001) will collect 
samples of millions of stars in our Galaxy as well as 
in our nearest neighbours with very accurate positions, 
proper motions, and trigonometric parallaxes 
which will dramatically improve the situation, and revolutionize
our knowledge of the Galaxy.

\begin{acknowledgements}
We wish to thank Volker Springel and Simon White who allowed us to use
data from their simulations; Antonaldo Diaferio and Ronald Drimmel 
for suggestions and many useful comments; Attilio Ferrari for his constant 
support to this project.  This work was initiated at the 
Astronomical Institute in Utrecht, which is gratefully acknowledged.  
P.R.F. wishes to thank the Astronomical Institute in Utrecht for hospitality 
during her visit. Partial financial support to this research comes from the
Italian Ministry of Research (MIUR) through the COFIN-2001 program,
and from the Netherlands Organization for Scientific Research (NWO)
and the Netherlands Research School for Astronomy (NOVA).
Finally, we thank the referee, Timothy Beers, for a careful reading of 
this manuscript and for his useful remarks.
\end{acknowledgements}



\begin{thebibliography}{}


\bibitem
[\protect\astroncite
{Beers \& Sommer-Larsen}{1995}]{pap1} 
Beers, T.C., \& Sommer-Larsen, J. 1995, ApJ, 96, 175 

\bibitem
[\protect\astroncite
{Beers et~al.}{2000}]{pap2}
Beers, T.C., Chiba, M., Yoshii, Y., et~al. 2000, ApJ, 119, 2866

\bibitem
[\protect\astroncite
{Beers et~al.}{2004}]{segue}
Beers, T.C., Allende P.C., Wilhelm, R., et~al. 2004, PASA, 21, 207



\bibitem
[\protect\astroncite
{Binney \& Merrifield}{1998}]{GA}
Binney, J., \& Merrifield, M. 1998, Galactic Astronomy. (Princeton: Princeton Univ. Press)

\bibitem
[\protect\astroncite
{Brook et~al.} {2003}]{brook}
Brook, C.B., Kawata, D., Gibson, B. K., \& Flynn, C. 2003, ApJ, 585, L125



\bibitem
[\protect\astroncite
{Chiba \& Beers}{2000}]{pap3}
Chiba, M., \& Beers, T.C. 2000, AJ, 119, 2843 (CB)

\bibitem
[\protect\astroncite
{Chiba \& Beers}{2001}]{CB2001}
Chiba, M., \& Beers, T.C. 2001, ApJ, 549, 325 


\bibitem
[\protect\astroncite
{Coil et~al.}{2004}]{coil}
Coil, A.L., Newman, J.A., Kaiser, N., et~al. 2004, ApJ, 617, 765


\bibitem
[\protect\astroncite
{Dehnen \& Binney}{1998}]{sun_lsr}
Dehnen, W., \& Binney, J.J. 1998, MNRAS, 298, 387

\bibitem
[\protect\astroncite
{Diemand et~al.}{2005}]{earth_halos}
Diemand, J., Moore, B., \& Stadel, J. 2005, Nature, 433, 389

\bibitem
[\protect\astroncite
{Eggen, Lynden-Bell \& Sandage}{1962}]{ELS}
Eggen, O.J., Lynden-Bell, D.A., \& Sandage, A.R. 1962, ApJ, 136, 748

\bibitem
[\protect\astroncite
{Ferguson et~al.}{2002}]{ferguson}
Ferguson, A.M.N., Irwin, M.J., Ibata, R.A., et al. 2002, AJ, 124, 1452

\bibitem
[\protect\astroncite
{Helmi et~al.}{1999}]{AHnat}
Helmi, A., White, S.D.M., de Zeeuw, P.T., \& Zhao, H.S. 1999, Nature, 402, 53 (HWdZZ)




\bibitem
[\protect\astroncite
{Helmi et~al.}{2002}]{ah_sim}
Helmi, A., White, S.D.M., \& Springel, V. 2002, Phys. Rev. D, 66, 063502

\bibitem
[\protect\astroncite
{Helmi et~al.} {2003}]{AH03} 
Helmi, A., White, S.D.M, \& Springel, V. 2003, MNRAS, 339, 834

\bibitem
[\protect\astroncite
{Helmi} {2004}]{AH04a}
Helmi, A. 2004a, MNRAS, 351, 643

\bibitem
[\protect\astroncite
{Helmi} {2004}]{AH04b}
Helmi, A. 2004b, ApJ, 610, L97

\bibitem
[\protect\astroncite
{Ibata et~al.}{1995}]{ibata}
Ibata, R.A., Gilmore, G., \& Irwin, M.J. 1995, MNRAS, 277, 781


\bibitem
[\protect\astroncite
{Johnston et~al.}{1996}]{JHB}
Johnston, K.V., Hernquist, L., \& Bolte, M. 1996, ApJ, 465, 278 

\bibitem
[\protect\astroncite
{Kerscher et~al.}{2000}]{CFcos}
Kerscher, M., Szapudi, I., \& Szalay, A.S. 2000, ApJ, 535, L13

\bibitem
[\protect\astroncite
{Law et~al.}{2005}]{law}
Law, D.R., Johnston, K.V., \& Majewski, S.R. 2005, ApJ, 619, 807

\bibitem
[\protect\astroncite
{Lemon et~al.}{2004}]{lemon}
Lemon, D.J., Wyse, R.F.G., Liske, J., Driver, S.P., \& Horne, K. 2004, MNRAS, 347, 1043

\bibitem
[\protect\astroncite
{Majewski, Munn, \& Hawley} {1996}]{MM96} 
Majewski, S.R., Munn, J.A., \& Hawley, S.L. 1996, ApJ, 456, L73

\bibitem
[\protect\astroncite
{Martin et~al.}{2004}]{martin}
Martin, N.F., Ibata, R.A., Bellazzini, M., et~al. 2004, MNRAS, 348, 12

\bibitem
[\protect\astroncite
{McLean et~al.} {2000}]{gsc}
McLean, B.J., Greene, G.R., Lattanzi, M. G., \& Pirenne, B. 2000, ASPC, 216, 145

\bibitem                                                           
[\protect\astroncite 
{Monet et~al.} {2003}]{usno}
Monet, D.G., Levine, S.E., Canzian, B., et~al. 2003, AJ, 125, 984


\bibitem
[\protect\astroncite
{Moore et~al.}{2001}]{moore_02}
Moore, B., Calcaneo-Roldan, C., Stadel, J., et al. 2001, Phys. Rev. D, 64, 063508


\bibitem
[\protect\astroncite
{Moore}{2001}]{moore_01}
Moore, B. 2001, AIPC, 586, 73

\bibitem
[\protect\astroncite
{Mullis et~al.}{2004}]{mullis}
Mullis, C.R., Henry, J.P., Gioia, I.M., et~al. 2004, ApJ, 617, 192

\bibitem
[\protect\astroncite
{Newberg et~al.} {2002}]{newberg}
Newberg, H.J., Yanny, B., Rockosi, C., et~al. 2002, ApJ, 569, 245 

\bibitem
[\protect\astroncite
{Nordstr\"om et~al.} {2004}]{nord}
Nordstr\"om, B., Mayor, M., Andersen, J., et~al. 2004, A\&A, 418, 989

\bibitem
[\protect\astroncite
{Perryman et~al.} {2001}]{gaia}
Perryman, M.A.C., de Boer, K.S., Gilmore, G., et~al. 2001, A\&A, 369, 339

\bibitem                                                            
[\protect\astroncite
{Platais et~al.} {1998}]{spm}
Platais, I., Girard, T.M., Kozhurina-Platais, V., et~al. 1998, AJ, 116, 2564 


%

\bibitem
[\protect\astroncite
{Reid \& Majewski} {1993}]{hZm}
Reid, I.N., \& Majewski S.R. 1993, ApJ, 409, 635

\bibitem
[\protect\astroncite
{Robertson et~al.} {2005}]{robertson}
Robertson, B., Bullock, J.S., Font, A.S., Johnston, K.V., \& Hernquist, L. 2005, astro-ph/0501398

\bibitem
[\protect\astroncite
{Robin et~al.}{1996}]{hZM}
Robin, A.C., Haywood, H., Cr\'ez\'e, M., Ojha, D. K., \& Bienaym\'e, O. 1996, A\&A, 305, 125

\bibitem
[\protect\astroncite
{Searle \& Zinn}{1978}]{SZ}
Searle, L., \& Zinn, R. 1978, ApJ, 225, 357

\bibitem
[\protect\astroncite
{Springel et~al.} {2001}]{springel}
Springel, V., White, S.D.M., Tormen, G., \& Kauffmann, G. 2001, MNRAS, 328, 726


\bibitem
[\protect\astroncite
{Steinmetz} {2003}]{rave}
Steinmetz, M. 2003, ASPC, 298, 381

\bibitem
[\protect\astroncite
{Zacharias et~al.} {2004}]{zach}
Zacharias, N., Urban, S.E., Zacharias, M.I., et~al. 2004, A\&A, 127, 3043


\bibitem
[\protect\astroncite
{Zheng et~al.}{1999}]{NGC5907}
Zheng, Z., Shang, Z., Su, H., et al. 1999, AJ, 117, 2757

\bibitem
[\protect\astroncite
{Zhao et~al.}{2005}]{zhao}
Zhao, H., Taylor, J.E., Silk, J., \& Hooper, D. 2005, astro-ph/0502049

\end{thebibliography}
\end{document}